\documentclass[11pt,thmsb,draft]{article}%
\usepackage{graphicx}
\usepackage{amsmath}%
\usepackage{amsfonts}%
\usepackage{amssymb}

\begin{document}

\title{Hidden Markov model segmentation\\of hydrological and enviromental time series}
\author{Ath. Kehagias}
\maketitle

\begin{abstract}
Motivated by Hubert's segmentation procedure \cite{Hub1997,Hub2000}, we
discuss the application of hidden Markov models (HMM) to the segmentation of
hydrological and enviromental time series. We use a HMM algorithm which
segments time series of several hundred terms in a few seconds and is
computationally feasible for even longer time series. The segmentation
algorithm computes the Maximum Likelihood segmentation by use of an
expectation / maximization iteration. We rigorously prove algorithm
convergence and use numerical experiments, involving temperature and river
discharge time series, to show that the algorithm usually converges to the
globally optimal segmentation. The relation of the proposed algorithm to
Hubert's segmentation procedure is also discussed.

\end{abstract}

\section{Introduction}

\label{sec01}

In this paper we discuss the following problem of \emph{time series
segmentation}:\emph{ }given a time series, divide it into two or
more\ \emph{segments} (i.e. blocks of contiguous data) such that each segment
is homogeneous, but contiguous segments are heterogeneous. Homogeneity /
heterogeneity is described in terms of some appropriate statistics of the
segments. The term \emph{change point detection }is also used to describe the problem.

Examples of this problem arise in a wide range of fields, including
engineering, computer science, biology and econometrics. The segmentation
problem is also relevant to hydrology and environmetrics. For instance, in
climate change studies it is often desirable to test a time series (such as
river flow, rainfall or temperature records) for one or more sudden changes of
its mean value.

The time series segmentation problem has been studied in the hydrological
literature. The reported approaches can be divided into two
categories:\ \emph{sequential} and \emph{nonsequential}. Sequential approaches
often involve \emph{intervention models}; see for example \cite{HipMac} and,
for a critique of intervention models, \cite{Rama1}.

Most of the nonsequential time segmentation work appearing in the hydrological
literature involves \emph{two} segments. In other words, the goal is to detect
the existence and estimate the location of a \emph{single} change point. A
classical early study of changes in the flow of Nile appears in \cite{Cobb1}.
Buishand's work \cite{Buis1,Buis2} is also often cited. For some case studies
see \cite{Hoppe1999,Kiel1998,Serv1997}. Bayesian approaches have recently
generated considerable interest
\cite{Per1999,Per2000a,Per2000b,Per2000c,Rama1}.

It appears that the \emph{multiple }change point problem has not been studied
as extensively. Hubert's segmentation procedure \cite{Hub1997,Hub2000} is an
important step in this direction. The goodness of a segmentation is evaluated
by the sum squared deviation of the data from the means of their respective
segments; in what follows we will use the term \emph{segmentation cost} for
this quantity. Given a time series, Hubert's procedure computes the
\emph{minimal cost }segmentation with $K=$2, 3, ... change points. The
procedure gradually increases $K$; for every value of $K$ the best
segmentation is computed; the procedure is terminated when differences in the
means of the obtained segments are no longer statistically significant (as
measured by Scheffe's contrast criterion \cite{Scheffe}). Hubert mentions that
this procedure can segment time series with several tens of terms but is ``...
unable at the present state to tackle series of much more than a hundred terms
...'' because of the combinatorial increase of computational burden
\cite{Hub2000}.

The work reported in this paper has been inspired by Hubert's procedure. Our
goal is to develop an algorithm which can locate multiple change points in
hydrological and/or environmental time series with several hundred terms or
more. To achieve this goal, we adapt some \emph{hidden Markov models (HMM)}
algorithms which have originally appeared in the speech recognition
literature. (A survey of the relevant literature is postponed to Section
\ref{sec0303}.) We introduce a HMM\ of hydrological and/or enviromental time
series with change points and describe an approximate \emph{Expectation /
Maximization (EM)\ algorithm} which produces a converging sequence of
segmentations. The algorithm also produces a sequence of estimates for the
HMM\ parameters. Time series of several hundred points can be segmented in a
few seconds (see Section \ref{sec04}), hence the algorithm can be used in an
interactive manner as an exploratory tool. Even for time series of several
thousand points the segmentation time is in the order of seconds.

This paper is organized as follows. In Section \ref{sec02} we review Hubert's
formulation of the time series segmentation problem. In Section \ref{sec03} we
formulate the segmentation problem in terms of hidden Markov models and
present a segmentation algorithm; also we compare the hidden Markov model
approach with that of Hubert. We present some segmentation experiments in
Section \ref{sec04}. In Section \ref{sec05} we summarize our results. Finally,
in the Appendix we present an alternative, non-HMM segmentation method, which
is more accurate but also slower.

\section{Time Series Segmentation as an Optimization Problem}

\label{sec02}

In this section we formulate time series segmentation as an optimization
problem. We follow Hubert's presentation, but we modify his notation.

Given a time series \ $\mathbf{x}$\ = ($x_{1}$, $x_{2}$, ... , $x_{T}$) and a
number $K$, a \emph{segmentation} is a sequence of times $\mathbf{t}$ =
$(t_{0}$, $t_{1}$, ... , $t_{K})$ which satisfy%
\begin{equation}
0=t_{0}<t_{1}<...<t_{K-1}<t_{K}=T.
\end{equation}
The intervals of integers $[t_{0}+1$, $t_{1}]$, $[t_{1}+1,...,t_{2}]$, ... ,
$[t_{K-1}+1,t_{K}]$ are the \emph{segments}; the times $t_{0}$, $t_{1}$, ... ,
$t_{K}$ are the \emph{change points}. $K$, the number of segments, is the
\emph{order} of the segmentation. The length of the $k$-th segment (for
$k=1,2,...,K$) is denoted by $T_{k}=t_{k}-t_{k-1}$. The following notation is
used for a given segmentation $\mathbf{t}$ = $(t_{0}$, $t_{1}$, ... , $t_{K}%
)$. For $k=1,2,...,K$, define%
\begin{equation}
\widehat{\mu}_{k}=\frac{\sum_{t=t_{k-1}+1}^{t_{k}}x_{t}}{T_{k}},\qquad
d_{k}=\sum_{t=t_{k-1}+1}^{t_{k}}\left(  x_{t}-\widehat{\mu}_{k}\right)  ^{2}.
\end{equation}
Define the \emph{cost }of segmentation $\mathbf{t}=(t_{0},...,t_{K})$ by%
\begin{equation}
D_{K}(\mathbf{t})=\sum_{k=1}^{K}d_{k}=\sum_{k=1}^{K}\sum_{t=t_{k-1}+1}^{t_{k}%
}\left(  x_{t}-\widehat{\mu}_{k}\right)  ^{2}. \label{eq99}%
\end{equation}
If $D_{K}$ has a small value, then the segments are homogeneous, i.e. the
$x_{t}$'s are close to $\widehat{\mu}_{k}$ for $k=1,2,...,K$ and for
$t=t_{k-1}+1,...,t_{k}$. Now we can define the best $K$-th order segmentation
$\widehat{\mathbf{t}}$ to be the one minimizing $D_{K}(\mathbf{t})$ and denote
the minimal cost by $\widehat{D}_{K}$ = $\widehat{D}_{K}(\widehat{\mathbf{t}%
})$. Note that for every $K$ we have $\widehat{D}_{K}\geq\widehat{D}_{K+1}$
\cite{Hub1997}. Also, there is only one segmentation $\mathbf{t}$ of order
$T$; in this case every time instant $t$ is a segment by itself and
$D_{T}(\mathbf{t})=0$.

It can be seen \cite{Hub1997} that the number of possible segmentations grows
exponentially with $T$. To efficiently search the set of all possible
segmentations, Hubert uses a \emph{branch-and-bound }approach. Even so, the
computational load increases excessively with $T$ and this approach is not
able currently (in 2000) to segment series of much more than a hundred terms
\cite{Hub2000}.

Minimization of $D_{K}$ can be achieved by several alternative (and faster
than branch-and-bound) algorithms. A \emph{dynamic programming} approach is
presented in the Appendix to obtain the globally minimum cost; this is
feasible for $T$ in the order of several hundreds and will be reported in
greater length in a future publication \cite{Keh01}. In this paper a different
approach is followed, which is based on HMM's.

\section{Hidden Markov Models}

\label{sec03}

We now present a HMM formulation of the time series segmentation problem.
HMM's have been used for runoff modeling \cite{LuBer1999} and the possibility
of using them for hydrological time series segmentation has been mentioned in
\cite{Per2000c} but, as far as the author knows, an actual implementation has
not been presented yet. On the other hand, we have already mentioned that
HMM's are used for segmentation of time series in several other fields (see
the discussion in Section \ref{sec0303}).

The term ``hidden Markov model'' is used to denote a broad class of stochastic
processes; here we use a particular and somewhat restricted species of HMM to
model a hydrological time series and present an approximate Expectation /
Maximization (EM) algorithm to perform \emph{Maximum Likelihood} (ML)
segmentation. In addition to the standard probabilistic interpretation of the
algorithm, a numerical optimization point of view is also possible and we use
the latter to prove the convergence of the algorithm. Finally we discuss
related algorithms and possible extensions.

\subsection{HMM's and Hydrological Time Series}

\label{sec0301}

We will use a pair of stochastic processes $(Z_{t},X_{t})$ to model a
hydrological time series with change points. We start by considering a simple example.

The annual flow of a river is denoted by $X_{t}$. We assume that, for the
years $t=1,2,...,t_{1}$, $X_{t}$ is a normally distributed random variable
with mean $\mu_{1}$ and standard deviation $\sigma$. In year $t_{1}$ a
\emph{transition} takes place and, for the years $t=t_{1}+1$, $t_{1}%
+2,...,t_{2}$, $X_{t}$ is normally distributed with mean $\mu_{2}$ and
standard deviation $\sigma$. This process continues with transitions taking
place in years $t_{2}$, $t_{3}$, ... , $t_{K-1}$. This process is illustrated
in Figure 1. We indicate the \emph{states} of the river flow by circles and
the possible transitions from state to state by arrows; note that the states
are \emph{unobservable}. We indicate the observable time series by the double
arrows emanating from the states.

\begin{center}
\textbf{Figure 1 to appear here}
\end{center}

The above mechanism can be modeled by a pair of stochastic processes
$(Z_{t},X_{t})$ (with $t=0,1,2,...$) defined as follows.

\begin{enumerate}
\item $Z_{t}$ , which is the \emph{state process}, is a finite state Markov
chain with $K$ states; it has initial probability vector $\pi$ and transition
probability matrix $P$. Hence, for any $T$, the joint probability function of
\ $Z_{0},Z_{1},...,Z_{T}$ is%
\begin{equation}
\Pr(Z_{1}=z_{1},Z_{2}=z_{2},...,Z_{T}=z_{T})=\pi_{z_{0}}\cdot P_{z_{0},z_{1}%
}\cdot P_{z_{1},z_{2}}\cdot...\cdot P_{z_{T-1},z_{Tn}}. \label{eq01}%
\end{equation}
For the specific example discussed above, it will also be true that:\ (a)
$\pi_{1}=1$, $\pi_{k}=0$ for $k=2,3,...,K$, (b)\ $P_{k,j}=0$ for $k=1,2,...,K$
and all $j$ other than $k$, $k+1$. The parameters of this process are $K$ and
$P$.

\item $X_{t}$ , which is the \emph{observation process}, is a sequence of
\emph{conditionally independent, }normally distributed random variables with
mean $\mu_{Z_{t}}$ and standard deviation $\sigma$. More precisely, for every
$t$, the joint probability density of \ $X_{1},X_{2},...,X_{t}$ conditioned on
$Z_{1},Z_{2},...,Z_{t}$ is%
\begin{equation}
f_{X_{1},X_{2},...,X_{t}|Z_{1},Z_{2},...,Z_{t}}(x_{1},x_{2},...,x_{t}%
|z_{1},z_{2},...,z_{t})=\prod_{i=1}^{n}e^{-\left(  x_{t}-\mu_{z_{t}}\right)
^{2}/2\sigma^{2}}. \label{eq02}%
\end{equation}
The parameters of this process are $\mu_{1}$, $\mu_{2}$, ... , $\mu_{K}$ and
$\sigma$. We will often use the notation $\mathbf{M}$ = [$\mu_{1}$, $\mu_{2}$,
... , $\mu_{K}$ ].
\end{enumerate}

The $(Z_{t},X_{t})$ pair is a HMM, in particular a \emph{left-to-right
continuous HMM} \cite{HMM04}. ``Left-to-right'' refers to the structure of
state transitions (as depicted in Figure 1) and ``continuous'' refers to the
fact that the observation process is continuous valued. The model parameters
are $K,$ $P,$ $\mathbf{M,}$ $\sigma$.

There is a one-to-one correspondence between state sequences $\mathbf{z}$
=\ ($z_{1}$, $z_{2}$, ... , $z_{T}$) and segmentations $\mathbf{t}$ = ($t_{0}%
$, $t_{1}$, ... , $t_{K^{\prime}}$). For example, given a particular
$\mathbf{z}$, we obtain the corresponding $\mathbf{t}$\ by locating the times
$t_{k}$ such that $z_{t_{k}}\neq z_{t_{k}+1}$, for $k=1,2,...,K^{\prime}-1$
(and setting $t_{0}=0$ and $t_{K^{\prime}}=T$). The postulated Markov chain
only allows left-to-right transitions, hence $K^{\prime}\leq K$, i.e. there
will be \emph{at most} $K\ $segments, and every segment will be uniquely
associated with a state.

The \emph{conditional likelihood }of a state sequence $\mathbf{z}$
(\emph{given }an observation sequence $\mathbf{x}$)\ is denoted by
\begin{equation}
L_{K,T}^{1}(\mathbf{z}|\mathbf{x};P,\mathbf{M,}\sigma)=L_{K,T}^{1}(z_{1}%
,z_{2},...,z_{T}|x_{1},x_{2},...,x_{T};P,\mathbf{M,}\sigma)\label{eq03}%
\end{equation}
and the \emph{joint likelihood }of a state sequence $\mathbf{z}$ \emph{and }
an observation sequence $\mathbf{x}$ is denoted by
\begin{equation}
L_{K,T}^{2}(\mathbf{z},\mathbf{x};P,\mathbf{M,\ }\sigma)=L_{K,T}^{2}%
(z_{1},z_{2},...,z_{T},x_{1},x_{2},...,x_{T};P,\mathbf{M,}\sigma).\label{eq04}%
\end{equation}
$L_{K,T}^{1}$ and $L_{K,T}^{2}$ are understood as functions of $\mathbf{z}$ =
$(z_{1},z_{2},...,z_{T})$; the observations $\mathbf{x}$ =\ ($x_{1}$, $x_{2}$,
... , $x_{T}$), the number of segments $K$, and the length of the time series
$T$, as well as the parameters $P,$ $\mathbf{M,}$ $\sigma$ are assumed
\emph{fixed}. In place of $T$ any $t$ can be used, to indicate the likelihood
of the subsequence $(z_{1},z_{2},...,z_{t})$ given $(x_{1},x_{2},...,x_{t})$
etc. For example, we can write
\begin{equation}
L_{K,t}^{2}(z_{1},z_{2},...,z_{t},x_{1},x_{2},...,x_{t};P,\mathbf{M,}%
\sigma).\label{eq06}%
\end{equation}
Note also that%
\begin{equation}
L_{K,T}^{1}=A\cdot L_{K,T}^{2}\label{eq05}%
\end{equation}
where $A$ is a quantity independent of $(z_{1},z_{2},...,z_{T})$. Finally,
from (\ref{eq01}), (\ref{eq02}) we have%
\begin{equation}
L_{K,T}^{2}(\mathbf{z},\mathbf{x};P,\mathbf{M,}\sigma)=\prod_{t=1}^{T}\left(
P_{z_{t-1},z_{t}}\cdot e^{-\left(  x_{t}-\mu_{z_{t}}\right)  ^{2}/2\sigma^{2}%
}\right)  ,\label{eq09}%
\end{equation}
where $z_{0}$ = 1, according to the previously stated assumption.

\subsection{The Segmentation Algorithm}

\label{sec0302}

The ML\ segmentation $\widehat{\mathbf{t}}$ can be obtained from the ML\ state
sequence $\widehat{\mathbf{z}}$ = ($\widehat{z}_{1}$, $\widehat{z}_{2}$, ... ,
$\widehat{z}_{T}$). Since state sequences are unobservable, we will estimate
$\widehat{\mathbf{z}}$ in terms of the observable sequence $\mathbf{x}$\ =
($x_{1}$, $x_{2}$, ... , $x_{T}$) and $\ $the \ parameters $K,$ $P,$
$\mathbf{M,}$ $\sigma$. Note that in practice $K,$ $P,$ $\mathbf{M,}$ $\sigma$
will also be unknown. Hence the computation of the maximum likelihood\ HMM
segmentation must be divided into two subtasks:\ (a)\ estimating the
HMM\ parameters and (b)\ computing the actual segmentation. We follow the
standard approach used in HMM\ problems: a parameter estimation phase is
followed by a time series segmentation phase and the process is repeated until
convergence. This is the Expectation / Maximization (EM) approach. First we
discuss estimation and segmentation in more detail; then we will return to a
discussion of the EM\ approach.

\subsubsection{Parameter Estimation}

\label{sec03021}

Suppose, for the time being, that a segmentation $\mathbf{t}$ = $(t_{0},t_{1}%
$,..., $t_{m})$ is given. A reasonable estimate of $\mathbf{M}$ = [$\mu_{1}$,
$\mu_{2}$, ... , $\mu_{K}$], \emph{dependent on the given segmentation}, is
(for $k=1,2,...,K$)%
\begin{equation}
\widehat{\mu}_{k}=\frac{\sum_{t=t_{k-1}+1}^{t_{k}}x_{t}}{T_{k}}. \label{eq21}%
\end{equation}
Similarly we could use the following \emph{segmentation-dependent }estimates
of $\sigma$ (for $k=1,2,...,K$)%
\begin{equation}
\widehat{\sigma}_{k}=\sqrt{\frac{\sum_{t=t_{k-1}+1}^{t_{k}}\left(
x_{t}-\widehat{\mu}_{k}\right)  ^{2}}{T_{k}-1}}. \label{eq23}%
\end{equation}
However, to maintain compatibility with Hubert's approach, we will use the
\emph{segmentation-independent }estimate%
\begin{equation}
\widehat{\sigma}=\sqrt{\frac{\sum_{t=1}^{T}\left(  x_{t}-\widehat{\mu}\right)
^{2}}{T-1}}=\sqrt{\frac{\sum_{k=1}^{K}\sum_{t=t_{k-1}+1}^{t_{k}}\left(
x_{t}-\widehat{\mu}\right)  ^{2}}{T-1}}. \label{eq22}%
\end{equation}
where
\[
\widehat{\mu}=\frac{\sum_{t=1}^{T}x_{t}}{T}%
\]

Let us now turn to the transition probability matrix $P$. In a left-to-right
HMM, for $k=1,2,...,K$ and all $j$ different from $k$ and $k+1$, we will have
$P_{k,j}=0$. Also, for $k=1,2,...,K-1$ we will have $P_{k,k+1}=1-P_{k,k}$.
Hence $P$ only has $K-1$ free parameters, namely $P_{1,1}$, $P_{2,2}$, ... ,
$P_{K-1,K-1}$. These could be estimated from the given segmentation. However,
in this paper we use a simpler approach. Namely, we assume%
\begin{equation}
P=\left[
\begin{array}
[c]{cccccc}%
p & 1-p & 0 & ... & 0 & 0\\
0 & p & 1-p & ... & 0 & 0\\
... & ... & ... & ... & ... & ...\\
0 & 0 & 0 & ... & p & 1-p\\
0 & 0 & 0 & 0 & 0 & 1
\end{array}
\right]  . \label{eq31}%
\end{equation}
Hence $P$ is determined in terms of a single parameter $p$, which will be
chosen a priori, rather than estimated. We have found by numerical
experimentation that the exact value of $p$ is not critical; in all the
examples of Section \ref{sec04}, the segmentation algorithm performs very well
using $p$ in the range [0.85,0.95].

Finally, we must make a choice regarding the number of segments $K$. We will
use Hubert's approach, and take a sequence of increasing values:\ $K=2$, $3$,
... until a value of $K$ is reached which yields statistically nonsignificant
segmentations (statistical significance is evaluated by Scheffe's contrast
criterion, \cite{Hub1997,Scheffe}).

\subsubsection{Segmentation}

\label{sec03022}

Given observations $\mathbf{x}=(x_{1},x_{2},...,x_{T})$ and assuming the
parameters $K$, $P$, $\mathbf{M}$\textbf{, }$\sigma$ to be \ known, the
\emph{Maximum Likelihood (ML)} state sequence is the $\widehat{\mathbf{z}}$ =
$(\widehat{z}_{1},\widehat{z}_{2},...,\widehat{z}_{T})$ which maximizes
$L_{K,T}^{1}(\mathbf{z}|\mathbf{x};P,\mathbf{M,}\sigma)$ as function of
$\mathbf{z}$. The ML\ segmentation $\widehat{\mathbf{t}}$ = ($\widehat{t}_{0}%
$, $\widehat{t}_{1}$, ... , $\widehat{t}_{K^{\prime}}$)\ is obtained from
$\widehat{\mathbf{z}}$. It will be seen in Section \ref{sec03024} that, under
certain circumstances, $\widehat{\mathbf{z}}\ $\ also minimizes the
segmentation cost $D_{K}$ defined in Section \ref{sec02}.

$\widehat{\mathbf{z}}$ $=$ $\ (\widehat{z}_{1},\widehat{z}_{2},...,\widehat
{z}_{T})$ can be found by the \emph{Viterbi algorithm }\cite{Vit01}, a
computationally efficient dynamic programming approach. In view of
(\ref{eq05}) we have
\begin{equation}
(\widehat{z}_{1},\widehat{z}_{2},...,\widehat{z}_{T})=\arg\max_{z_{1}%
,z_{2},...,z_{T}}L_{K,T}^{2}(z_{1},z_{2},...,z_{T},x_{1},x_{2},...,x_{T}%
;P,\mathbf{M,}\sigma). \label{eq11}%
\end{equation}
Now, for $t=1,2,...,T$ $\ $and\emph{ } $k=1,2,...,K$ define%
\begin{equation}
q_{k,t}=\max_{z_{1},z_{2},...,z_{t-1}}L_{K,t}^{2}(z_{1},z_{2},...,z_{t-1}%
,k,x_{1},x_{2},...,x_{t};P,\mathbf{M,}\sigma) \label{eq012}%
\end{equation}
It can be shown by standard dynamic programming arguments \cite{DP01} that
both $\widehat{\mathbf{z}}$ $=$ $\ (\widehat{z}_{1}$, $\widehat{z}_{2}$,
$...$, $\widehat{z}_{T})$ and the $q_{k,t}$'s of (\ref{eq012}) can be computed
recursively as follows.

\begin{center}
\rule{6in}{0.02in}

\textbf{Viterbi Algorithm}
\end{center}

\begin{itemize}
\item[ ] \textbf{Input}: The time series $x_{1},x_{2},...,x_{T}$ ; the
parameters $K$, $P$, $\mathbf{M}$ and $\sigma$.

\item[ ] \textbf{Forward Recursion}

\item[ ] Set $q_{1,0}=1$, $q_{2,0}=q_{3,0}=...=q_{k,0}=0$.

\item[ ] For $t=1,2,...,T$ $\ $\emph{ }

\begin{itemize}
\item[ ] For $k=1,2,...,K$
\begin{align*}
\text{ }q_{k,t}  &  =\max_{1\leq j\leq K}\left(  q_{j,t-1}\cdot P_{j,k}\cdot
e^{-\left(  x_{t}-\mu_{k}\right)  ^{2}/2\sigma^{2}}\right) \\
r_{k,t}  &  =\arg\max_{1\leq j\leq K}\left(  q_{j,t-1}\cdot P_{j,k}\cdot
e^{-\left(  x_{t}-\mu_{k}\right)  ^{2}/2\sigma^{2}}\right)  .
\end{align*}

\item[ ] End
\end{itemize}

\item[ ] End

\item[ ] \textbf{Backtracking}

\item[ ] $\widehat{L}_{K,T}^{2}=\max_{1\leq k\leq K}\left(  q_{k,T}\right)  $

\item[ ] $\widehat{z}_{T}=\arg\max_{1\leq k\leq K}\left(  q_{k,T}\right)  $.

\item[ ] For $t=T,T-1,...,2$%
\[
\widehat{z}_{t-1}=r_{\widehat{z}_{t},t}.
\]

\item[ ] End
\end{itemize}

\begin{center}
\rule{6in}{0.02in}
\end{center}

Upon completion of the forward recursion, $\widehat{L}_{K,T}^{2}$, the maximum
value of $L_{K,T}^{2}$, is obtained. The backtracking phase produces the state
sequence which maximizes $L_{K,T}^{2}$ (and hence also $L_{K,T}^{1}$ ).
Execution time is of order O($T\cdot K^{2}$) which is \emph{linear} (rather
than exponential)\ in the length of the time series $T$. This makes the
algorithm computationally feasible even for long time series. For more details
on the Viterbi algorithm see \cite{Vit01}.

\subsubsection{Combined Parameter Estimation and Segmenation}

\label{sec03023}

Parameter estimation and segmentation can be combined in an algorithm which
maximizes the likelihood viewed as a function of \emph{both }the state
sequence $\mathbf{z}$ = $(z_{1},z_{2}$,..., $z_{T})$ and the parameters
$\mathbf{M}$. The algorithm presented below is an iterative \emph{Expectation
/ Maximization} (EM)\ algorithm \cite{EM01}\ which produces a converging
sequence of segmentations.

\begin{center}
\rule{6in}{0.02in}

\textbf{HMM\ Segmentation Algorithm}
\end{center}

\begin{itemize}
\item[ ] \textbf{Input:}\ The time series $\mathbf{x=(}x_{1},x_{2},...,x_{T})$
; the parameters $K$, $P$; a termination variable $\varepsilon$.

\item[ ] Choose randomly a state sequence $\widehat{\mathbf{z}}^{(0)}$\textbf{
= (}$z_{1}^{(0)},z_{1}^{(0)}$,..., $z_{T}^{(0)}$).

\item[ ] Compute $\widehat{\sigma}$ from $\ $(\ref{eq22}).

\item[ ] For $i=1,2,...$ $\ $\emph{ }

\begin{itemize}
\item[ ] Compute $\mathbf{t}^{(i)}\ $from $\widehat{\mathbf{z}}^{(i-1)}$.

\item[ ] Compute $\widehat{\mathbf{M}}^{(i)}$ from $\mathbf{t}^{(i)}\ $and
(\ref{eq21}).

\item[ ] Compute $\widehat{\mathbf{z}}^{(i)}$ by the Viterbi algorithm using
$\mathbf{x}$, $K$, $P$, $\mathbf{M}^{(i)}$ and $\widehat{\sigma}$.

\item[ ] If
$\vert$%
$L_{K,T}^{2}(\widehat{\mathbf{z}}^{(i)},\mathbf{x};P$, $\widehat{\mathbf{M}%
}^{(i)}$,$\widehat{\sigma})$ $-$ $L_{K,T}^{2}(\widehat{\mathbf{z}}%
^{(i-1)},\mathbf{x};P$, $\widehat{\mathbf{M}}^{(i-1)}$,$\widehat{\sigma
})|<\varepsilon$.

\begin{itemize}
\item[ ] $\widehat{\mathbf{z}}$ = $\widehat{\mathbf{z}}^{(i)}$.

\item[ ] Exit the loop
\end{itemize}

\item[ ] EndIf
\end{itemize}

\item[ ] End
\end{itemize}

\begin{center}
\rule{6in}{0.02in}
\end{center}

In Section \ref{sec03024} we will show that the above algorithm is a very
close approximation to an EM\ algorithm and that, under certain conditions,
every iteration increases the likelihood function. In all the examples
presented in Section \ref{sec04} the algorithm converges to the \emph{global}
maximum with very few iterations (typically 3 or 4). In other words, the outer
loop of the algorithm is executed only a few times; in each execution we
perform a parameter reestimation according to (\ref{eq21}) (with execution
time O($T$))\ and a segmentation by the Viterbi algorithm (with execution time
O($T\cdot K^{2}$)). Hence the total execution time for a fixed $K$ value is
O($T\cdot K^{2}$).

For a complete segmentation procedure the above algorithm is run for a
sequence of increasing values $K=2,3,...$ $.$ First the algorithm is used to
obtain the ML\ segmentation of order $K=$2; the difference of the means of the
two segments is tested for statistical significance by the Scheffe criterion
(for details see \cite{Hub1997} and \cite{Scheffe}). If the difference is not
significant, then it is concluded that the entire time series consists of a
single segment. If the difference is significant, the algorithm is run with
$K=3$ and the Scheffe test is applied to the resulting segments. The process
is continued until, for some value of $K$, a segmentation is obtained which
fails the Scheffe test (or until we reach $K=T$, an unlikely case). 

The use of Scheffe's contrast criterion to determine the true value of $K$ is
somewhat problematic. This point is discussed in some detail in \cite{Hub1997}%
. Many methods for the determination of $K$ have been proposed in the
literature, but none of these completely resolves the problem. In cases of
doubt, a pragmatic approach would be to use human judgement to evaluate
segmentations with different $K$'s. In the case of hydrological and
environmental time series which involve a rather small number of segments,
this is relatively easy. The short execution time of the segmentation
algorithm favors this approach, since experimentation in an ``interactive''
mode is feasible.

\subsubsection{Convergence}

\label{sec03024}

The goal of this section is to show that, for a fixed $K$, every iteration of
the HMM\ segmentation algorithm increases the likelihood; since the likelihood
is bounded above by one, this also implies that the algorithm converges$.$

Two approaches can be used. The first approach is based upon the probabilistic
interpretation of the algorithm; since this is a routinely applied analysis of
EM\ algorithms, it will be presented only in outline. In the second approach,
the \ segmentation algorithm is viewed from a numerical optimization point of
view and convergence is proved without using any probabilistic assumptions;
furthermore this approach shows clearly the connection of our segmentation
algorithm to Hubert's procedure.

\noindent\textbf{Probabilistic Approach}. As explained in \cite{EM01}, the
basic ingredient of the EM\ family of algorithms is the iterative application
of an expectation step followed by a likelihood maximization step. In our case
the expectation step consists in estimating $\mathbf{M}^{(i)}\ $by
(\ref{eq21}) and the maximization step consists in finding $\mathbf{z}^{(i)}$
by the Viterbi algorithm.

While the Viterbi algorithm computes exactly the global maximum of the
likelihood (viewed as a function of $\mathbf{z}$ only!), the estimation step
used in this paper is approximate. The exact step would involve computing
estimates of $\widehat{\mu}_{1}$, $\widehat{\mu}_{2}$, $...$, $\widehat{\mu
}_{K}$ for every possible segmentation and then combining these estimates in a
sum weighted by the respective probability of each segmentation (a similar
approach should be used for $\sigma$, using the estimates of (\ref{eq23})).
This approach is used in \cite{HMM01} and elsewhere; while it is
computationally more expensive than the approach used here, it is still
viable. At any rate, in most cases the two approaches yield very similar results.

If it is assumed that the estimate of (\ref{eq21}) is a close approximation to
the maximum likelihood estimate of $\mathbf{M}$, then convergence can be
established by a standard EM\ argument presented in \cite{EM01,HMM03} and
several other places. This argument shows that a certain cross entropy
$Q(\mathbf{z}^{(i)},\mathbf{z}^{(i-1)})$ is decreased by every iteration of an
EM algorithm. Since $Q$ is always nonnegative, it must converge to a
nonnegative number, and this suffices for the algorithm to terminate.
Furthermore, by relating $Q(\mathbf{z}^{(i)},\mathbf{z}^{(i-1)})$ to the
likelihood, it can be shown that the sequence $L_{K,T}(\mathbf{z}^{(i)})$ is
monotonically increasing.

\noindent\textbf{Numerical Approach}. In what follows we will consider $K$,
$P$, $\mathbf{x}$\textbf{, }$\mathbf{\sigma}$ to be fixed. We will denote the
set of all possible state sequences by $\Phi$ and the set of all state
sequences with $K$ transitions by $\Phi_{K}$; we will also use the standard
notation $\mathbf{R}^{K}$ for the set of all $K$-dimensional real vectors.

Taking the negative logarithm of (\ref{eq09}) we obtain%
\begin{equation}
-\log\left[  L_{K,T}^{2}(\mathbf{z},\mathbf{x};P,\mathbf{M,}\sigma)\right]
=-\sum_{t=1}^{T}\log\left(  P_{z_{t-1},z_{t}}\right)  +\sum_{t=1}%
^{T}\frac{\left(  x_{t}-\mu_{z_{t}}\right)  ^{2}}{2\sigma^{2}}.\label{eq41}%
\end{equation}
We define $\phi(\mathbf{z})=$ ``number of times $z_{t-1}\neq z_{t}$''; in
other words, $\phi(\mathbf{z})$ is the number of transitions in the state
sequence $\mathbf{z}$. If we limit ourselves to state sequences $\mathbf{z}%
\in\Phi_{K}$, then obviously $\phi(\mathbf{z})=K$. Now, for all $\mathbf{z}%
\in\Phi_{K}$, (\ref{eq41})\ becomes%
\begin{align}
-\log\left[  L_{K,T}^{2}(\mathbf{z},\mathbf{x};P,\mathbf{M,}\sigma)\right]
&  =-\left(  (T-\phi(\mathbf{z}))\cdot\log\left(  p\right)  +\phi
(\mathbf{z})\cdot\log\left(  1-p\right)  \right)  +\sum_{t=1}^{T}\frac{\left(
x_{t}-\mu_{z_{t}}\right)  ^{2}}{2\sigma^{2}}\label{eq42}\\
&  =-\left(  (T-K)\cdot\log\left(  p\right)  +K\cdot\log\left(  1-p\right)
\right)  +\sum_{t=1}^{T}\frac{\left(  x_{t}-\mu_{z_{t}}\right)  ^{2}}%
{2\sigma^{2}}\Rightarrow\label{eq43}\\
&  =C(T,K,P)+\sum_{t=1}^{T}\frac{\left(  x_{t}-\mu_{z_{t}}\right)  ^{2}%
}{2\sigma^{2}}.\label{eq44}%
\end{align}
where $C(T,K,P)$ = $-\left[  (T-K)\cdot\log\left(  p\right)  +K\cdot
\log\left(  1-p\right)  \right]  $. Now we define the function \
\begin{equation}
J(\mathbf{z,M})=\sum_{t=1}^{T}\left(  x_{t}-\mu_{z_{t}}\right)  ^{2}%
\end{equation}
and note that
\begin{equation}
J(\mathbf{z,M})=-2\sigma^{2}\cdot\left(  \log\left[  L_{K,T}^{2}%
(\mathbf{z},\mathbf{x};P,\mathbf{M,}\sigma)\right]  +C(T,K,P)\right)
.\label{eq53}%
\end{equation}
Note that, for simplicity of notation, we write $J(\mathbf{z,M})$ as a
function only of $\mathbf{z,M}$; the quantities $T$, $K$, $P$, $\mathbf{x}%
$\textbf{, }$\mathbf{\sigma}$ can be considered fixed.

Now consider a run of the segmentation algorithm which produces a sequence
$\mathbf{z}^{(0)}$, $\mathbf{z}^{(1)}$, $\mathbf{z}^{(2)}...\ $ ,
$\mathbf{z}^{(i)}$, ...$\ $ . \emph{Suppose that for every }$s$\emph{ we have}
$\mathbf{z}^{(i)}\in\Phi_{K}$. By the reestimation formula for $\mathbf{M}%
^{(i)}$ we will have for every $s$:%
\begin{equation}
\forall\mathbf{M}\in\mathbf{R}^{K}:J(\mathbf{z}^{(i-1)};\mathbf{M}^{(i)})\leq
J(\mathbf{z}^{(i-1)};\mathbf{M}). \label{eq51}%
\end{equation}
Furthermore, note that the Viterbi algorithm yields the global maximum of the
likelihood \emph{as a function of }$\mathbf{z}$. Hence, from (\ref{eq53}) and
the reestimation formula for $\mathbf{z}^{(i)}$ we will have for every $i$:
\begin{equation}
\forall\mathbf{z}\in\Phi_{K}:J(\mathbf{z}^{(i)};\mathbf{M}^{(s)})\leq
J(\mathbf{z};\mathbf{M}^{(i)}). \label{eq52}%
\end{equation}
Now, using first (\ref{eq52}) and then (\ref{eq51}), we obtain%
\begin{equation}
J(\mathbf{z}^{(i)};\mathbf{M}^{(i)})\leq J(\mathbf{z}^{(i-1)};\mathbf{M}%
^{(i)})\leq J(\mathbf{z}^{(i-1)};\mathbf{M}^{(i-1)}) \label{eq55}%
\end{equation}
and, from (\ref{eq55}) and (\ref{eq53}),%
\begin{equation}
L_{K,T}^{2}(\mathbf{z}^{(i)},\mathbf{x};P,\mathbf{M}^{(i)}\mathbf{,}%
\sigma)\geq L_{K,T}^{2}(\mathbf{z}^{(i-1)},\mathbf{x};P,\mathbf{M}%
^{(i-1)}\mathbf{,}\sigma) \label{eq56}%
\end{equation}

Hence, \emph{if for every }$i$\emph{ we have} $\mathbf{z}^{(i)}\in\Phi_{K}$,
then the sequence $\left\{  L_{K,T}^{2}(\mathbf{z}^{(i)},\mathbf{x}%
;P,\mathbf{M}^{(i)}\mathbf{,}\sigma)\right\}  _{i=0}^{\infty}$ is increasing;
since it is also bounded from above by one, it must converge. It follows that
the HMM\ segmentation algorithm produces a sequence of segmentations with
increasing and convergent likelihood; from convergence of the likelihood we
also conclude that the algorithm will eventually terminate. Furthermore, if
$\mathbf{t}^{(i)}$ is the segmentation obtained from $\mathbf{z}^{(i)}$ is
easy to check that
\begin{equation}
D_{K}(\mathbf{t}^{(i)})=J(\mathbf{z}^{(i)};\mathbf{M}^{(i)}).\label{eq57}%
\end{equation}
From (\ref{eq51}), (\ref{eq57}) follows that \emph{Hubert's segmentation cost
is decreased in every iteration }of the HMM\ segmentation algorithm.

For the above analysis to hold, we have required that $z^{(i)}\in\Phi_{K}$ for
every $i$. This condition is easy to check; it is usually satisfied in
practice; and it can be \emph{enforced} by choosing the parameter $p$ to be
not too close to 1 (if $p\simeq1$, then the cost of state transitions is very
high and transitions are avoided).

One way to interpret the above analysis is the following: using an appropriate
value of $p$, the segmentation algorithm presented here becomes an iterative,
approximate way to find Hubert's optimal segmentation. The approximation is
usually very good, as will be seen in Section \ref{sec04}. This interpretation
is completely nonprobabilistic and does not depend on the use of the hidden
Markov model.

\noindent\textbf{Computational Issues}. We must also mention that succesful
implementation of the Viterbi algorithm requires a normalization of the
$q_{k,t}$'s to avoid numerical underflow; alternatively one can work with the
logarithms of the the $q_{k,t}$'s and perform additions rather than multiplications.

\subsection{Discussion and Extensions}

\label{sec0303}

An extensive mathematical, statistical and engineering literature covers both
the theoretical and applied aspects of HMM's. The reader can use
\cite{HMM01,HMM04} as starting points for a broader overview of the subject.
EM-like algorithms for HMM's were introduced in
\cite{HMM02a,HMM02b,HMM02c,HMM03}. The EM\ family of algorithms was introduced
in great generality in \cite{EM01}; work on HMM's also appears in the
econometrics \cite{Hamil1,Krolz1}, as well as in the biological \cite{Baldi}
literature. These references are merely starting points; the literature is
very extensive.

As already mentioned, the EM\ segmentation algorithm used here is a variation
of algorithms which are well-established \ in the field of speech recognition;
for example see \cite{HMM06,HMM07}. Taking into account the extensive
HMM\ literature, as well as various ideas reported in the hydrological
literature, the algorithm of Section \ref{sec03024} can be extended in several directions.

\begin{enumerate}
\item The assumption that the observations are normally distributed is not
essential. Other forms of probability density can be used in (\ref{eq09}).
Similarly, by a simple modification of (\ref{eq09}) the algorithm can handle
vector valued observations.

\item A basic idea of the algorithm is that each segment must be
\emph{homogeneous}. Assuming that the observations within a segment are
generated independently and normally, segment homogeneity is evaluated by the
deviation of $x_{t_{k-1}+1},x_{t_{k-1+2}},...,x_{t_{k}}$ from the segment mean
$\widehat{\mu}_{k}$. But alternative assumptions can be used. For example,
assume that the observations are generated by an autoreggressive mechanism,
i.e. that, for $t=t_{k-1}+1,t_{k-1}+2,...,t_{k}$ and $k=1,2,...,K$, we have%
\begin{equation}
x_{t}=a_{0,k}+a_{1,k}x_{t-1}+a_{2,k}x_{t-2}+...+a_{l,k}x_{t-l}+\epsilon_{t}
\label{eq71}%
\end{equation}
(where $\epsilon_{t}$ is a white noise term). The segmentation algortithm can
be used within this framework. In this case the reestimation phase computes
the AR$\;$coefficients $a_{1,k}$, $\ a_{2,k}$, $...$ , $a_{l,k}$, which can be
estimated from $x_{t_{k-1}+1}$, $x_{t_{k-1+2}}$, $...$ , $x_{t_{k}}$ using a
least squares fitting algorithm. This approach is used in Section
\ref{sec0403} to fit a HMM\ autoregressive model to global temperature data.

\item Similarly, it may be assumed that the observations are generated by a
polynomial regression of the form (for $t=t_{k-1}+1,t_{k-1}+2,...,t_{k}$ and
$k=1,2,...,K$)%
\begin{equation}
x_{t}=a_{0,k}+a_{1,k}\cdot(t-t_{k-1})+...+a_{l,k}\cdot(t-t_{k-1})^{l}%
+\epsilon_{t} \label{eq72}%
\end{equation}
where $\epsilon_{t}$ is a noise term. Again, the coefficients $a_{0,k}$,
$\ a_{1,k}$, $...$ , $a_{l,k}$ can be computed at every reestimation phase by
a least squares fitting algorithm. Additional constraints can be used to
enforce continuity across segments. In the case of 1st order polynomials there
are only two coefficients, $a_{0,k}$, $\ a_{1,k}$, which are determined by the
continuity assumptions; the iterative reestimation of the change points can
still be performed. This case may be of interest for detection of trends.

\item It has been mentioned in Section \ref{sec03021} that $P$ can also be
reestimated in every iteration of the EM\ algorithm. Preserving the
left-to-right structure implies that for $k=1,2,...,K$ and for all $j$
different from $k$ and $k+1$, we have $P_{k,j}=0$; furthermore, for
$k=1,2,...,K-1$ we have $P_{k,k+1}=1-P_{k,k}$. The $P_{k,k}$ parameters can be
estimated by $\widehat{P}_{k,k}=\frac{T_{k}}{T_{k}+1}$. However, some
preliminary experiments indicate that this approach does not yield improved segmentations.

\item On the other hand, the treatment of the state transition can be modified
in a more substantial manner by dropping the left-to-right assumption. In the
current model each state of the Markov chain corresponds to a single segment
and, because of the left-to-right structure, it is visited at most once. An
alternate approach would be to assign some physical significance to the
states. For instance, states could be chosen to correspond to climate regimes
such as ``dry'', ``wet'' etc. In this case a state could be visited more than
once. This approach allows the choice of models which incorporate expert
knowledge about the evolution of climate regimes. On the other hand, if the
left-to-right structure is dropped, the number of free parameters in the $P$
matrix increases. These parameters could be estimated (conditional on a
particular state sequence)\ by
\begin{equation}
\widehat{P}_{kj}=\frac{\text{no. of times that }z_{t}=k\text{ and }z_{t+1}%
=j}{\text{no. of times that }z_{t}=k}.
\end{equation}
The enhancements of arbitrary transition structure and transition probability
estimation are easily accommodated by our algorithm.
\end{enumerate}

\section{Experiments}

\label{sec04}

In this section we evaluate the segmentation algorithm by numerical
experiments. The first experiment involves an annual river discharge time
series which contains 86 points. The second example involves the reconstructed
annual mean global temperature time series and contains 282 points. Both of
these examples involve segmentation by minimization of total deviation from
segment means. The third example again involves the annual mean global
temperature time series, but performs segmentation by minimization of
autoregressive prediction error. The fourth example involves artificially
generated time series with up to 1500 points.

\subsection{Annual Discharge of the Senegal \ River}

\label{sec0401}

In this experiment we use the time series of the Senegal river annual
discharge data, measured at the Bakel station for the years 1903-1988. The
length of the time series is 86. The same data set has been used by Hubert
\cite{Hub1997,Hub2000}. The goal is to find the segmentation which is optimal
with respect to total deviation from the segment means, has the highest
possible order and is statistically significant according to Scheffe's criterion.

We run the segmentation algorithm for increasing values of $K$. In the
experiments reported here we have always used $p=0.9$ (similar results are
\ obtained for other values of $p$ in the interval [0.85, 0.95]. For every
value of $K$, convergence is achieved by the 3rd or 4th iteration of the
algorithm. The optimal segmentations are presented in Table 1. The
segmentations which were validated by the Scheffe criterion appear in bold letters.

\begin{center}%
\begin{tabular}
[c]{|l|l|l|l|l|l|l|l|}\hline
$K$ & \multicolumn{7}{|l|}{Segment Boundaries (Change Points)}\\\hline
1 & \textbf{1902} & \textbf{1988} &  &  &  &  & \\\hline
2 & \textbf{1902} & \textbf{1967} & \textbf{1988} &  &  &  & \\\hline
3 & \textbf{1902} & \textbf{1949} & \textbf{1967} & \textbf{1988} &  &  &
\\\hline
4 & \textbf{1902} & \textbf{1917} & \textbf{1953} & \textbf{1967} &
\textbf{1988} &  & \\\hline
5 & \textbf{1902} & \textbf{1921} & \textbf{1936} & \textbf{1949} &
\textbf{1967} & \textbf{1988} & \\\hline
6 & 1902 & 1921 & 1936 & 1949 & 1967 & 1971 & 1988\\\hline
\end{tabular}

\textbf{Table 1}
\end{center}

Hence it can be seen that the optimal and statistically significant
segmentation is that of order 5, i.e. the segments are [1903,1921],
[1922,1936], [1937,1949], [1950,1967], [1967,1988]. That this is the globally
optimal \ segmentation, has been shown by Hubert in \cite{Hub1997,Hub2000}
using his exact segmentation procedure. A plot of the time series, indicating
the 5 segments and the respective means appears in Figure 2.

\begin{center}
\textbf{Figure 2 to appear here}
\end{center}

We have verified that the HMM\ algorithm finds the globally optimal
segmentation for all values of $K$ (as listed in Table 1). We performed this
verification by use of the exact dynamic programming algorithm presented in
the Appendix. The conclusion is that, in this experiment, the
HMM\ segmentation algorithm finds the optimal segmentations considerably
faster than the exact algorithm. Specifically, running the entire experiment
(i.e. obtaining the HMM\ segmentations of \emph{all} orders) with a
MATLAB\ implementation of the HMM\ segmentation algorithm took 1.1 sec on a
Pentium III 1 GHz personal computer; we expect that a FORTRAN\ or
C\ implementation would take about 10\% to 20\% of this time.

\subsection{Annual Mean Global Temperature}

\label{sec0402}

In this experiment we use the time series of \ annual mean global temperature
for the years 1700 -- 1981. Only the temperatures for the period 1902 -- 1981
come from actual measurements; the remaining temperatures were
\emph{reconstructed} according to a procedure described in \cite{Mann1999} and
also at the Internet address
\texttt{http://www.ngdc.noaa.gov/paleo/ei/ei\_intro.html}. \ The length of the
time series is 282. The goal is again to find the segmentation which is
optimal with respect to total deviation from the segment-means, has the
highest possible order and is statistically significant according to Scheffe's criterion.

We run the segmentation algorithm for $K=2,3,...,6$, using $p=0.9$.
Convergence takes place in 4 iterations or less. The optimal segmentations are
presented in Table 2. The segmentations which were validated by Scheffe's
criterion appear in bold letters.

\begin{center}%
\begin{tabular}
[c]{|l|l|l|l|l|l|l|l|}\hline
$K$ & \multicolumn{7}{|l|}{Segment Boundaries (Change Points)}\\\hline
1 & \textbf{1700} & \textbf{1981} &  &  &  &  & \\\hline
2 & \textbf{1700} & \textbf{1930} & \textbf{1981} &  &  &  & \\\hline
3 & \textbf{1700} & \textbf{1812} & \textbf{1930} & \textbf{1981} &  &  &
\\\hline
4 & \textbf{1700} & \textbf{1720} & \textbf{1812} & \textbf{1930} &
\textbf{1981} &  & \\\hline
5 & 1700 & 1720 & 1812 & 1926 & 1935 & 1981 & \\\hline
6 & 1700 & 1720 & 1812 & 1926 & 1934 & 1977 & 1981\\\hline
\end{tabular}

\textbf{Table 2}
\end{center}

Hence it can be seen that the optimal and statistically significant
segmentation is of order 4, i.e. the segments are [1700,1720], [1721,1812],
[1813,1930], [1931,1981]. A plot of the time series, indicating the 4 segments
and the respective means appears in Figure 3.

\begin{center}
\textbf{Figure 3 to appear here}
\end{center}

The \emph{total} execution time for the experiment (i.e. to obtain optimal
segmentations of all orders) is 2.97 sec. The segmentations of Table 2 are the
globally optimal ones, as we have verified using the dynamic programming
segmentation algorithm.

\subsection{Annual Mean Global Temperature with AR model}

\label{sec0403}

In this experiment we again use the annual mean global temperature time
series, but now we assume that it is generated by a \emph{switching regression
}HMM. Specifically, we assume a model of the form%

\begin{equation}
x_{t}=a_{0,k}+a_{1,k}x_{t-1}+a_{2,k}x_{t-2}+a_{3,k}x_{t-3}+\epsilon_{t}
\label{eq46}%
\end{equation}
where the parameters $a_{0,k}$, $a_{1,k}$, $a_{2,k}$, $a_{3,k}$ are specific
to the $k$-th state of the underlying Markovian process. Given a particular
segmentation, these parameters can be estimated by a least squares fitting
algorithm. Hence the segmentation algorithm can be modified to obtain the
optimal segmentation with respect to the model of (\ref{eq46}).

Once again we run the segmentation algorithm for $K=2,3,...,6$, using $p=0.9$.
The optimal segmentations thus obtained are presented in Table 3.

\begin{center}%
\begin{tabular}
[c]{|l|l|l|l|l|l|l|l|}\hline
$K$ & \multicolumn{7}{|l|}{Segment Boundaries (Change Points)}\\\hline
1 & \textbf{1700} & \textbf{1981} &  &  &  &  & \\\hline
2 & \textbf{1700} & \textbf{1926} & \textbf{1981} &  &  &  & \\\hline
3 & \textbf{1700} & \textbf{1833} & \textbf{1926} & \textbf{1981} &  &  &
\\\hline
4 & \textbf{1700} & \textbf{1769} & \textbf{1833} & \textbf{1926} &
\textbf{1981} &  & \\\hline
5 & 1700 & 1769 & 1833 & 1895 & 1926 & 1981 & \\\hline
6 & 1700 & 1769 & 1825 & 1877 & 1904 & 1926 & 1981\\\hline
\end{tabular}

\textbf{Table 3}
\end{center}

In this case segment validation is not performed by the Scheffe criterion;
instead we use a prediction error correlation criterion. This indicates the
maximum statistically significant number of segments is $K$=4 and the segments
are [1700,1769], [1770,1833], [1834,1926], [1927,1981]. A plot of the time
series, indicating the 4 segments and the respective autoregressions appears
in Figure 3.

\begin{center}
\textbf{Figure 4 to appear here}
\end{center}

Recall that the segments obtained by means-based segmentation are [1700,1720],
[1721, 1812], [1813, 1930], [1931, 1981]. This seems to be in reasonable
agreement with the AR-based segmentation, excepting the discrepancy of 1720
and 1769. From a numerical point of view, there is no a priori reason to
expect that the AR-based segmentation and means-based segmentation should give
the same results. The fact that the two segmentations are in reasonable
agreement, supports the hypothesis that actual climate changes have occurred
approximately at the transition times indicated by both segmentation methods.

Finally, let us note that the \emph{total} execution time for the experiment
(i.e. to obtain optimal segmentations of every order) is 3.07 sec and that the
segmentations of Table 3 are the globally optimal ones, as we have verified
using the dynamic programming segmentation algorithm.

\subsection{Artificial Time Series}

\label{sec0404}

The goal of the final experiment is to investigate the scaling properties of
the algorithm, specifically the scaling of execution time with respect to time
series length $T$ and the scaling of accuracy with respect to noise in the
observations. To obtain better control over these factors, artificial time
series are used, which have been generated by the following mechanism.

The time series are generated by a 5-th order HMM. Every time series is
generated by running the HMM from state no.1 until state no.5. Hence, every
time series involves 5 state transitions and, for the purposes of this
experiment, this is assumed to be known a priori. On the other hand, it can be
seen that the length of the time series is variable. With a slight change of
notation, in this section $T$ will denote the \emph{expected} length of the
time series, which can be controlled by choice of the probability $p$. The
values of $p$ were chosen to generate time series of average lengths 200, 250,
500, 750, 1000, 1250, 1500.

The observations are generated by a normal distribution with mean $\mu_{k}$
($k$= 1, 2, ..., 5) and standard deviation $\sigma$. In all experiments the
values $\mu_{1}$= $\mu_{3}$= $\mu_{5}$= 1, $\mu_{2}$= $\mu_{4}$= $-1$ were
used. Several values of $\sigma$ were used, namely $\sigma$= 0.00, 0.10, 0.20,
0.30, 0.50, 0.75, 1.00, 1.25, 1.50, 1.75, 2.00.

For each combination of $T$ and $\sigma,$ 20 time series were generated and
the HMM\ segmentation algorithm was run on each one. For each run two
quantities were computed: $c$, accuracy of segmentation, and $T_{e}$,
execution time. Segmentation accuracy is computed by the formula%
\[
c=\frac{\sum_{t=1}^{T}\mathbf{1}(z_{t}=\widehat{z}_{t})}{T}%
\]
where the indicator function $\mathbf{1}(z_{t}=\widehat{z}_{t})$ is equal to 1
when $z_{t}=\widehat{z}_{t}$ and equal to 0 otherwise.

From these data two tables are compiled. Table 4 lists $T_{e}$ (in
seconds)\ as a function of $T$ (i.e. $T_{e}$ is averaged over all time series
of the same $T$). Table 5 lists average segmentation accuracy $c$ as a
function of $T$ and $\sigma$ (i.e. $c$ is averaged over the 20 time series
with the same $T$ and $\sigma$). As expected, segmentation accuracy is
generally a decreasing function of $\sigma$.

\begin{center}%
\begin{tabular}
[c]{|l|l|l|l|l|l|l|l|}\hline
$T$ & 200 & 250 & 500 & 750 & 1000 & 1250 & 1500\\\hline
$T_{e}$ & 0.193 & 0.249 & 0.585 & 1.024 & 1.845 & 3.026 & 4.60\\\hline
\end{tabular}

\textbf{Table 4. }Average execution time $T_{e}$ (in seconds)\ as a function
of average time series length $T$.

\bigskip

\medskip%
\begin{tabular}
[c]{|l|lllllll|}\hline
$T$ & 200 & \multicolumn{1}{|l}{250} & \multicolumn{1}{|l}{500} &
\multicolumn{1}{|l}{750} & \multicolumn{1}{|l}{1000} &
\multicolumn{1}{|l}{1250} & \multicolumn{1}{|l|}{1500}\\\hline
$\sigma$ & \multicolumn{7}{|c|}{$c$}\\\hline
0.00 & 1.0000 & 1.0000 & 1.0000 & 0.9692 & 1.0000 & 1.0000 &
0.9902\\\cline{1-1}%
0.10 & 1.0000 & 1.0000 & 1.0000 & 0.9814 & 1.0000 & 1.0000 &
1.0000\\\cline{1-1}%
0.20 & 1.0000 & 0.9806 & 1.0000 & 1.0000 & 1.0000 & 0.9716 &
1.0000\\\cline{1-1}%
0.30 & 1.0000 & 1.0000 & 0.9999 & 0.9792 & 1.0000 & 0.9807 &
1.0000\\\cline{1-1}%
0.50 & 0.9989 & 0.9993 & 0.9994 & 0.9997 & 1.0000 & 0.9997 &
1.0000\\\cline{1-1}%
0.75 & 0.9945 & 0.9979 & 0.9663 & 0.9521 & 0.9988 & 0.9992 &
0.9991\\\cline{1-1}%
1.00 & 0.9881 & 0.9880 & 0.9863 & 0.9974 & 0.9517 & 0.9981 &
0.9711\\\cline{1-1}%
1.25 & 0.9778 & 0.9710 & 0.9762 & 0.9924 & 0.9965 & 0.9843 &
0.9781\\\cline{1-1}%
1.50 & 0.9561 & 0.9701 & 0.9874 & 0.9341 & 0.9507 & 0.9362 &
0.9956\\\cline{1-1}%
1.75 & 0.9337 & 0.8985 & 0.9494 & 0.9341 & 0.9708 & 0.9272 &
0.9942\\\cline{1-1}%
2.00 & 0.8628 & 0.8617 & 0.8255 & 0.9141 & 0.8600 & 0.9523 & 0.8297\\\hline
\end{tabular}
\end{center}

\noindent\textbf{Table 5. }Average classif. accuracy $c$ as a function of
average time series length $T$ and noise level $\sigma$.

\section{Conclusion}

\label{sec05}

In this paper we have used hidden Markov models to represent hydrological and
enviromental time series with multiple change points. Inspired by Hubert's
pioneering work and by methods of speech recognition, we have presented a fast
iterative segmentation algorithm which belongs to the EM\ family. The quality
of a particular segmentation is evaluated by the deviation from segment means,
but extensions involving autoregressive HMM's, trend-generating HMM's etc. can
also be used. Because execution time is O($T\cdot K^{2}$), our algorithm can
be used to explore various possible segmentations in an interactive manner. We
have presented a convergence analysis which shows that under appropriate
conditions every iteration of our algorithm increases the likelihood of the
resulting segmentation. Furthermore, numerical experiments (involving river
flow and global temperature time series) indicate that the algorithm can be
expected to converge to the \emph{ globally }optimal segmentation.

\appendix     

\section{Appendix: A Dynamic Programming Segmentation Algorithm}

\label{seca}

In this appendix we present an alternative time series segmentation algorithm
which, unlike the HMM\ algorithm, is\emph{ guaranteed }to produce the
\emph{globally optimal }segmentation of a time series. This superior
performance, however, is obtained at the price of longer execution time.
Still, the algorithm is computationally viable for time series of several
hundred terms. We describe the algorithm briefly here; a more detailed report
appears in \cite{Keh01}.

\subsection{A General Segmentation Cost}

\label{seca01}

A \emph{generalization} of the time series segmentation problem discussed in
previous sections is the following. Given a time series $\mathbf{x}=(x_{1}$,
$x_{2}$, ... , $x_{T})$ and a fixed $K$, find a sequence of times $\mathbf{t}$
= $(t_{0}$, $t_{1}$, ... , $t_{K})$ which satisfies $0=t_{0}<t_{1}<$ ...
$<t_{K-1}<t_{K}$ = $T$, and minimizes%
\begin{equation}
J_{K}(\mathbf{t})=\sum_{k=1}^{K}f_{k}(t_{k-1},t_{k};\mathbf{x}). \label{eq101}%
\end{equation}
$J_{K}(\mathbf{t})$ consists of a sum of terms $f_{k}(t_{k-1},t_{k}%
;\mathbf{x})$. For example, Hubert's cost function can be obtained by setting%
\begin{equation}
f_{K}(s,t;\mathbf{x})=\sum_{\tau=s+1}^{t}\left(  x_{\tau}-\frac{\sum
_{\tau=s+1}^{t}x_{\tau}}{t-s}\right)  ^{2}. \label{eq102}%
\end{equation}
Hence Hubert's segmentation cost (\ref{eq99}) is a special case of
(\ref{eq101}).

Similarly, consider \emph{autoregressive} models of the form%
\begin{equation}
x_{t}=u_{t}A_{k}+\epsilon_{t}, \label{eq103}%
\end{equation}
where $t=t_{k-1}+1$, $t_{k-1}+2$, ... , $t_{k}$) and $u_{t}=[1$, $x_{t-1}$,
$x_{t-2}$, $...$, $x_{t-l}]$, $A_{k}=[a_{k,1}$, $a_{k,2}$, $...$,
$a_{k,l}]^{\prime}$ (the $^{\prime}$ denotes transpose of a matrix). Then we
can set
\begin{equation}
f_{K}(s,t;\mathbf{x})=\sum_{\tau=s+1}^{t}\left(  x_{\tau}-u_{\tau}%
A_{k}\right)  ^{2}. \label{eq105}%
\end{equation}
Then the segmentation cost becomes%
\begin{equation}
J_{K}(\mathbf{t})=\sum_{\tau=s}^{t}\epsilon_{\tau}^{2}=\sum_{k=1}^{K}%
\sum_{t=t_{k-1}+1}^{t_{k}}\left(  x_{t}-u_{t}A_{k}\right)  ^{2}. \label{eq104}%
\end{equation}
The $a_{k,1}$, $a_{k,2}$, $...$, $a_{k,l}$ (elements of $A_{k}$) are unknown,
but can be determined by least squares fitting on $x_{t_{k-1}+1}$,
$x_{t_{k-1}+2}$, ... , $x_{t_{k}}$. A similar formulation can be used for
regressive models of the form $x_{t}=u_{t}A_{k}+\epsilon_{t}$ where $A_{k}$ =
$[a_{k,0}$, $a_{k,1}$, $...$, $a_{k,l}]^{\prime}$, $u_{t}$ = $[1$,
$(t-t_{k-1})$, $(t-t_{k-1})^{2}$, $...$, $(t-t_{k-1})^{l}]$. Hence we see that
(\ref{eq101})\ is sufficiently general to subsume\ many cost functions of
practical interest.

\subsection{Dynamic Programming Segmentation Algorithm}

\label{seca02}

The following dynamic programming algorithm can be used to minimize
(\ref{eq101}); it has been presented in \cite{Auger} and applies to very
general versions of the time series segmentation problem.

\begin{center}
\rule{6in}{0.02in}

\textbf{Dynamic Programming Segmentation Algorithm}
\end{center}

\begin{itemize}
\item[ ] \textbf{Input:}\ The time series $\mathbf{x=(}x_{1},x_{2},...,x_{T}%
)$; a termination number $K$.

\item[ ] \textbf{Initialization}

\item[ ] For $t=1,2,...,T$ $\ $\emph{ }

\begin{itemize}
\item[ ] For $s=1,2,...,t$ $\ $\emph{ }

\begin{itemize}
\item[ ] $d_{s,t}=f_{K}(s-1,t;\mathbf{x})$
\end{itemize}

\item[ ] End

\item[ ] $c_{t,0}=d_{1,t}$
\end{itemize}

\item[ ] End

\item[ ] \textbf{Minimization}

\item[ ] For $k=1,2,...,K$

\begin{itemize}
\item[ ] For $t=k,k+1,...,T$

\begin{itemize}
\item[ ] For $s=0,1,...,t-1$

\begin{itemize}
\item[ ] $e_{s}=c_{s,k-1}+d_{s+1,t}$
\end{itemize}

\item[ ] End

\item[ ] $c_{t,k}=\min_{1\leq s\leq t}\left(  e_{s}\right)  $

\item[ ] $z_{t,k}=\arg\min_{1\leq s\leq t}\left(  e_{s}\right)  $
\end{itemize}

\item[ ] End
\end{itemize}

\item[ ] End

\item[ ] \textbf{Backtracking}

\item[ ] For $k=1,2,...,K$

\begin{itemize}
\item[ ] $\widehat{t}_{k,k}=T$

\item[ ] For $n=k-1,k-2,...,1$

\begin{itemize}
\item[ ] $\widehat{t}_{n,k}=z_{\widehat{t}_{n+1,k},n}$
\end{itemize}

\item[ ] End

\item[ ] $\widehat{t}_{0,k}=0$
\end{itemize}

\item[ ] End
\end{itemize}

\rule{6in}{0.02in}

On termination, the dynamic programming segmentation algorithm has computed
\begin{equation}
c_{T,k}=\min_{\mathbf{t=(}t_{0},t_{1},...,t_{k})}J_{k}(\mathbf{t})
\end{equation}
for $k=1,2,...,K$; in other words it has recursively solved a \emph{sequence
}of minimization problems. For $k=1,2,...,K$, the optimal segmentation
$\widehat{\mathbf{t}}_{k}$ = ($t_{0,k}$, $t_{1,k}$, ... , $t_{k,k}$) has been
obtained by backtracking.

The recursive minimization is performed in the second part of the algorithm;
it is seen that computation time is O($K\cdot T^{2}$). This is not as good as
the O($K^{2}\cdot T$) obtained by the HMM algorithm (note that usually $K$
is$\ $significantly less than $T$), but is still computationally viable for
$T\ \ $in the order of a few hundreds. The backtracking part of the algorithm
has execution time O($K^{2}$).

However, in many cases the computationally most expensive part of the
algorithm is the initialization phase, i.e. the computation of $d_{s,t}$. This
involves O($T^{2}$) computations of $d_{s,t}=f_{K}(s-1,t;\mathbf{x})$ and can
increase the computation cost by one or more orders of magnitude. For example,
if we apply the algorithm to detect changes in the mean, then
\begin{equation}
d_{s,t}=f_{K}(s-1,t;\mathbf{x})=\sum_{\tau=s}^{t}\left(  x_{\tau}%
-\frac{\sum_{\tau=s}^{t}x_{\tau}}{t-s+1}\right)  ^{2} \label{eq111}%
\end{equation}
which involves $t-s+1$ addittions; if (\ref{eq111}) is used in the
initialization phase, then this phase requires O($T^{3}$) computations and
this severely limits computational viability to relatively short time series.

Hence, to enhance the computational viability of the dynamic programming
segmentation algorithm, it is necessary to find efficient ways to perform the
initialization phase. In the next two sections, we will deal with this
question for two specific forms of $f_{K}(s,t;\mathbf{x})$: the first form
pertains to the computation of means and the second to the computation of
regressions and autoregressions.\ 

\subsection{Fast Computation of Means}

\label{seca03}

The computation of means can be performed recursively, as will now be shown.
For $t=1,2,...,T$, $s=1,2,...,t-1$, we must compute%
\begin{equation}
M_{s,t}=\sum_{\tau=s}^{t}x_{\tau}\text{,\qquad}d_{s,t}=f_{k}(s-1,t;\mathbf{x}%
)=\sum_{\tau=s}^{t}\left(  x_{\tau}-\frac{M_{s,t}}{t-s+1}\right)  ^{2}.
\label{eq121}%
\end{equation}
For $t=1,2,...,T$, $s=1,2,...,t$, define the following additional quantities:%
\begin{equation}
p_{s,t}=\frac{\sum_{\tau=s}^{t}x_{\tau}}{\sum_{\tau=s}^{t}1},\qquad
q_{s,t}=p_{s+1,t}-p_{s,t}. \label{eq122}%
\end{equation}
Then we have%
\begin{equation}
d_{s,t}=\sum_{\tau=s}^{t}(x_{\tau}-p_{s,t})^{2}=(x_{s}-p_{s,t})^{2}+\sum
_{\tau=s+1}^{t}(x_{\tau}-p_{s,t})^{2} \label{eq141}%
\end{equation}
and%
\begin{align}
\sum_{\tau=s+1}^{t}(x_{\tau}-p_{s,t})^{2}  &  =\sum_{\tau=s+1}^{t}(x_{\tau
}-p_{s+1,t}-p_{s+1,t}-p_{s,t})^{2}\nonumber\\
&  =\sum_{\tau=s+1}^{t}(x_{\tau}-p_{s+1,t})^{2}+\sum_{\tau=s+1}^{t}%
(p_{s+1,t}-p_{s,t})^{2}+2\cdot\sum_{\tau=s+1}^{t}(x_{\tau}-p_{s+1,t}%
)(p_{s+1,t}-p_{s,t})\nonumber\\
&  =d_{s+1,t}+(t-s)\cdot(q_{s,t})^{2}+2\cdot(p_{s+1,t}-p_{s,t})\cdot\left(
\sum_{\tau=s+1}^{t}x_{\tau}-(t-s)p_{s+1,t}\right)  \Rightarrow\nonumber\\
\sum_{\tau=s+1}^{t}(x_{\tau}-p_{s,t})^{2}  &  =d_{s+1,t}+(t-s)\cdot
(q_{s,t})^{2} \label{eq142}%
\end{align}
From (\ref{eq141}), (\ref{eq142}) follows that (for $t=1,2,...,T$,
$s=1,2,...,t-1$)%
\begin{equation}
d_{s,t}=d_{s+1,t}+(t-s)\cdot\left(  q_{s,t}\right)  ^{2}+(x_{s}-p_{s,t})^{2}.
\label{eq123}%
\end{equation}
The above computations can be implemented in time O($T^{2}$) by the following algorithm.

\begin{center}
\rule{6in}{0.02in}

\textbf{Recursive Computation of }$d_{s,t}$
\end{center}

\begin{itemize}
\item[ ] For $t=1,2,...,T$ $\ $\emph{ }

\begin{itemize}
\item[ ] $M_{t,t}=x_{t}$

\item[ ] $p_{t,t}=M_{t,t}$

\item[ ] For $s=t-1,t-2,...,1$ $\ $\emph{ }

\begin{itemize}
\item[ ] $M_{s,t}=x_{s}+M_{s+1,t}$

\item[ ] $p_{s,t}=\frac{M_{s,t}}{t-s+1}$
\end{itemize}

\item[ ] End
\end{itemize}

\item[ ] End

\item[ ] For $t=1,2,...,T$ $\ $\emph{ }

\begin{itemize}
\item[ ] For $s=1,2,..,t-1$

\begin{itemize}
\item[ ] $q_{s,t}=(p_{s+1,t}-p_{s,t})$
\end{itemize}

\item[ ] End
\end{itemize}

\item[ ] End

\item[ ] For $t=1,2,...,T$ $\ $\emph{ }

\begin{itemize}
\item[ ] $d_{t,t}=0$

\item[ ] For $s=t-1,t-,2,...,1$ $\ $\emph{ }

\begin{itemize}
\item[ ] $d_{s,t}=d_{s+1,t}+(t-s)\cdot\left(  q_{s,t}\right)  ^{2}%
+(x_{s}-p_{s,t})^{2}.$
\end{itemize}

\item[ ] End
\end{itemize}

\item[ ] End
\end{itemize}

\rule{6in}{0.02in}

Hence, if the above code replaces the initialization phase of the dynamic
programming algorithm in Section \ref{seca02}, we obtain an O($K\cdot T^{2}$)
implementation of the entire algorithm. In other words, we obtain an algorithm
which, given a time series of length $T$, computes the global minimum of
Hubert's segmentation cost (for all segmentations of orders $K=1,2,3,...,T$)
in time O($K\cdot T^{2}$)

\subsection{Fast Computation of Regression Coefficients}

\label{seca04}

Consider now autoregressive models described by (\ref{eq103}). As already
mentioned, in this case we have%
\begin{equation}
f_{k}(t_{k-1},t_{k};\mathbf{x})=\sum_{t=t_{k-1}+1}^{t_{k}}\left(  x_{t}%
-u_{t}A_{k}\right)  ^{2}.\label{eq131}%
\end{equation}
Hence $d_{s,t}=f_{k}(s-1,t;\mathbf{x})$ is given by%
\begin{equation}
d_{s,t}=\sum_{\tau=s}^{t}\left(  x_{\tau}-u_{\tau}A(s,t)\right)
^{2}.\label{eq132}%
\end{equation}
where $u_{t}=[1$, $x_{t-1}$, $x_{t-2}$, ... , $x_{t-l}]$ and $A(s,t)$ is
obtained by solving the least squares equation
\begin{equation}
A(s,t)=\left(  U(s,t)^{\prime}\cdot U(s,t)\right)  ^{-1}\cdot U(s,t)^{\prime
}\cdot X(s,t)\label{eq133}%
\end{equation}
with%
\begin{equation}
X(s,t)=\left[
\begin{array}
[c]{l}%
x_{s}\\
x_{s+1}\\
...\\
x_{t}%
\end{array}
\right]  \qquad\text{ and}\qquad\text{ }U(s,t)=\left[
\begin{array}
[c]{l}%
u_{s}\\
u_{s+1}\\
...\\
u_{t}%
\end{array}
\right]  .\label{eq134}%
\end{equation}
Note that to solve (\ref{eq133}) the matrix multiplications $U(s,t)^{\prime
}\cdot U(s,t)$, $U(s,t)^{\prime}\cdot X(s,t)$ must be performed. For
$t=1,2,...,T$, $s=1,2,...,t$, these multiplications require O($T^{5}$) time.
However, the solution of (\ref{eq133}) \ can be approximated by a fast
recursive algorithm reported in \cite{Graupe}. Choose some small number
$\delta$ and set%
\begin{equation}
P_{0}=\frac{1}{\delta}\cdot I\label{eq135}%
\end{equation}
(where $I$ is the $(l+1)\times(l+1)$ unit matrix). Then, consider the
following recursion for $s=1,2,...,T$ and $t=s+1,...,T$:%
\begin{align}
u_{t} &  =[1,x_{t-1},x_{t-2},...,x_{t-l}],\label{eq136a}\\
n &  =t-s,\label{eq136b}\\
P_{n} &  =P_{n-1}-P_{n-1}\cdot u_{t}^{\prime}\cdot u_{t}\cdot P_{n-1}%
\cdot\frac{1}{1+u_{t}\cdot P_{n-1}\cdot u_{t}^{\prime}},\label{eq136c}\\
\widehat{A}(s,t) &  =\widehat{A}(s,t-1)+P_{n}\cdot u_{t}^{\prime}\cdot\left(
x_{t}-u_{t}\cdot\widehat{A}(s,t-1)\right)  .\label{eq136d}%
\end{align}
Using the arguments of \cite{Graupe} for a fixed $s$ and increasing $t$ it can
be shown that $\widehat{A}(s,t)$ converges \emph{very quickly} to $A(s,t)$,
the true solution of (\ref{eq133}). Furthermore, the computations of
(\ref{eq136a})-(\ref{eq136d}) can be implemented in time O($T^{2}$). Hence,
for the case of autoregressive models, the $d_{s,t}$ computation can be
programmed as follows.

\begin{center}
\rule{6in}{0.02in}

\textbf{Recursive Computation of }$d_{s,t}$
\end{center}

\begin{itemize}
\item[ ] For $s=1,2,...,T$ $\ $\emph{ }

\begin{itemize}
\item[ ] $P_{0}$=$\frac{1}{\delta}\cdot I$

\item[ ] Initialize $\widehat{A}(s,s)$ randomly

\item[ ] $d_{s,s}$=0

\item[ ] For $t=s+1,s+2,...,T$

\begin{itemize}
\item[ ] $u_{t}=[1,x_{t-1},x_{t-2},...,x_{t-l}]$

\item[ ] $n=t-s$

\item[ ] $P_{n}$ = $P_{n-1}-P_{n-1}\cdot u_{t}^{\prime}\cdot u_{t}\cdot
P_{n-1}\cdot\frac{1}{1+u_{t}\cdot P_{n-1}\cdot u_{t}^{\prime}}$

\item[ ] $\widehat{A}(s,t)$ = $\widehat{A}(s,t-1)+P_{n}\cdot u_{t}^{\prime
}\cdot\left(  x_{t}-u_{t}\cdot\widehat{A}(s,t-1)\right)  $

\item[ ] $d_{s,t}=d_{s,t-1}+\left(  x_{t}-u_{t}\cdot\widehat{A}(s,t)\right)
^{2}$
\end{itemize}

\item[ ] End
\end{itemize}

\item[ ] End
\end{itemize}

\rule{6in}{0.02in}

Hence, if the above code replaces the initialization phase of the dynamic
programming segmentation algorithm in Section \ref{seca02}, we have an
O($K\cdot T^{2}$) implementation of the entire algorithm for autoregressive
models. A similar modification is possible for regressive models of the form
(\ref{eq103}).

\begin{center}
\end{center}

\end{document}